\documentclass[lettersize,journal]{IEEEtran}
\usepackage{amsmath,amsfonts}
\usepackage{algorithm}
\usepackage{array}
\usepackage[caption=false,font=normalsize,labelfont=sf,textfont=sf]{subfig}
\usepackage{textcomp}
\usepackage{stfloats}
\usepackage{url}
\usepackage{verbatim}
\usepackage{graphicx}
\usepackage{cite}

\usepackage{upgreek}
\usepackage{kitcolors}
\usepackage{amsmath}
\usepackage{mleftright}

\usepackage{mathtools}
\usepackage{hyperref}
\usepackage{siunitx}
\usepackage{tikz}
\usepackage{bm}
\usepackage{algpseudocode}
\usepackage[normalem]{ulem}

\definecolor{cb-1}{HTML}{4477AA}
\definecolor{cb-2}{HTML}{EE6677}
\definecolor{cb-3}{HTML}{228833}
\definecolor{cb-4}{HTML}{CCBB44}
\definecolor{cb-5}{HTML}{66CCEE}
\definecolor{cb-6}{HTML}{AA3377}
\definecolor{cb-7}{HTML}{BBBBBB}

\DeclareSIUnit{\belmilliwatt}{Bm}
\DeclareSIUnit{\dBm}{\deci\belmilliwatt}
\algnewcommand{\LineComment}[1]{\State \(\triangleright\) #1}
\begin{document}

\title{End-to-End Learning of Probabilistic Constellation Shaping through Importance Sampling}

\author{Shrinivas Chimmalgi, Laurent Schmalen,~\IEEEmembership{Fellow,~IEEE,} and Vahid Aref
\thanks{Shrinivas Chimmalgi and Laurent Schmalen are with the Communications Engineering Lab (CEL), Karlsruhe Institute of Technology (KIT), Germany (e-mail:shrinivas.chimmalgi@kit.edu; laurent.schmalen@kit.edu).}
\thanks{Vahid Aref is with Nokia, Magirusstr. 8, 70469 Stuttgart, Germany (e-mail:vahid.aref@nokia.com).}
\thanks{This work has received funding from the European Research Council (ERC) under the European Union's Horizon 2020 research and innovation programme (grant agreement No. 101001899).}
}

\markboth{Journal of \LaTeX\ Class Files,~Vol.~14, No.~8, August~2021}%
{Shell \MakeLowercase{\textit{et al.}}: A Sample Article Using IEEEtran.cls for IEEE Journals}

\maketitle

\begin{abstract}

 Probabilistic constellation shaping enables easy rate adaption and has been proven to reduce the gap to Shannon capacity. Constellation point probabilities are optimized to maximize either the mutual information or the bit-wise mutual information. The optimization problem is however challenging even for simple channel models. While autoencoder-based machine learning has been applied successfully to solve this problem~\cite{Aref22}, it requires manual computation of additional terms for the gradient which is an error-prone task. In this work, we present novel loss functions for autoencoder-based learning of probabilistic constellation shaping for coded modulation systems using automatic differentiation and importance sampling. We show analytically that our proposed approach also uses exact gradients of the constellation point probabilities for the optimization. In simulations, our results closely match the results from~\cite{Aref22} for the additive white Gaussian noise channel and a simplified model of the intensity-modulation direct-detection channel.
\end{abstract}


\section{Introduction}
\IEEEPARstart{P}{robabilistic} constellation shaping (PCS) is a candidate for improving the performance of fiber-optic communication systems~\cite{Dar2014a,Geller2016,Amari2019,Cho2019}. PCS allows to finely adapt the information rate and has also been demonstrated to have an improved tolerance to fiber nonlinearities~\cite{Fehenberger2016,Amari2019,Pilori2019,Fu2021}. The constellation point probabilities can be optimized for maximizing either the mutual information (MI) or the bit-wise mutual information (BMI). End-to-end (E2E) machine learning using autoencoders has been successfully applied to this challenging optimization problem~\cite{Boecherer2017,Stark2019JointLO,Aoudia2020JointLO,Aref22}. E2E approaches usually use gradient ascent methods for the optimization by leveraging powerful automatic differentiation tools such as autodiff~\cite{autodiff}. It was shown in~\cite{Aref22} that the gradients of MI or BMI in terms of the constellation point probabilities calculated using automatic differentiation can be incorrect. Additional terms must be manually added to the gradients calculated by automatic differentiation to ensure correct gradient-based optimization. The method was shown to work well for the additive white Gaussian noise (AWGN) channel~\cite{Aref22} and also for the nonlinear optical fiber channel~\cite{Neskorniuk2022a}.

While the method from~\cite{Aref22} has been proven to work well even on nonlinear channels, there is room for improvement on two fronts. In the case of strong shaping, depending on the type of distribution approximation, using constant composition distribution matching~\cite[Sec. 2.5]{CIT-111} for sampling a short symbol sequence may result in the low probability constellation points not being sampled. This can negatively impact learning for those points. The second more important aspect is the need to manually compute additional terms for the gradient which is an error-prone task. In this paper we address both these issues by proposing new cost functions for maximizing either the MI or the BMI. The cost functions allow the use of relatively short symbol sequences and the gradients of the cost functions in terms of surrogates of the constellation point probabilities can be calculated correctly using automatic differentiation alone.

\section{Probabilistic Shaping by Learning Weights}
\begin{figure*}
    \centering 
    \includegraphics[width=0.8\textwidth]{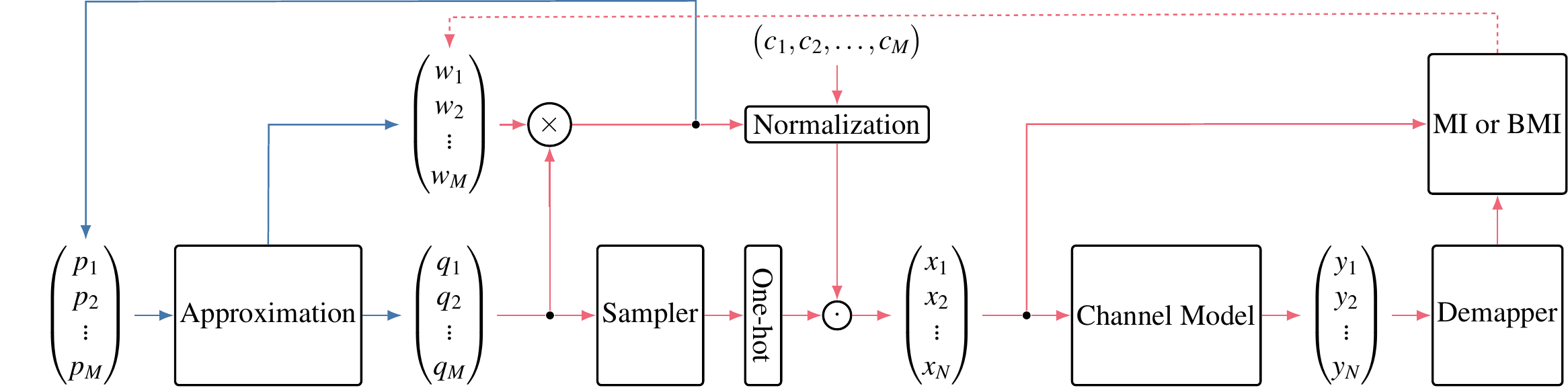}
    \captionsetup{width=0.8\textwidth}
    \caption{Block diagram of the proposed end-to-end learning system. The arrow directions mark the order of operations. Operations marked with \textcolor{cb-2}{\rule[0.5ex]{1em}{1pt}} are executed once per batch while those marked with \textcolor{cb-1}{\rule[0.5ex]{1em}{1pt}} are executed once per epoch. Dashed line indicates the optimization update of $\mathcal{W}_M$ based on MI or BMI per batch.}
    \label{fig:block_diagram}    
\end{figure*}

In Fig.~\ref{fig:block_diagram}, we show the block diagram of our proposed autoencoder. Let $\mathcal{C}_M=\{c_1,c_2,\ldots,c_M\}$ denote the set of constellation points with corresponding probabilities $\mathcal{P}_M=\{p_1,p_2,\ldots,p_M\}$. We define a related set of ``sampling'' probabilities $\mathcal{Q}_M=\{q_1,q_2,\ldots,q_M\}$ with $q_m>0$. We now define weights $\mathcal{W}_M=\{w_1,w_2,\ldots,w_M\}$ as $w_m\coloneqq p_m/q_m$. If symbol metric decoding is used at the receiver, the achievable information rate (AIR) is lower bounded by~\cite[Eq. 34]{Arnold2006},
\begin{equation}
\begin{aligned}
I(X;Y)&\geq  -\sum_{{m=1}}^{M}q_m w_m \log_2(q_m)-\sum_{{m=1}}^{M}q_m w_m \log_2(w_m)\\
 &+\sum_{{m=1}}^{M} \int_y q_m w_m p_{Y|X}\left(y|\tilde{c}_m\right)\log_2\left(Q_{X|Y}(\tilde{c}_m|y) \right) \mathrm{d}y,   
\end{aligned}    
\label{eq:MI_w}
\end{equation}

where $y$ is the continuous channel output, $p_{Y|X}$ is the channel law, $\tilde{c}_m$ are the normalized constellation points and $Q_{X|Y}$ is the posterior probability (output) of our autoencoder receiver. The \texttt{Approximation}, \texttt{Sampler}, \texttt{Normalization}, \texttt{Channel Model} and \texttt{Demapper} blocks can differ based on the optimization target. For a channel with an average power constraint on the inputs, the normalization is given by, $\tilde{c}_m=c_m/\sqrt{\sum_{{m=1}}^{M}p_m|c_m|^2}$. When $Q_{X|Y}$ is not derived from the channel law $p_{Y|X}$, \eqref{eq:MI_w} gives us a lower bound on the AIR, which is achievable by a maximum-likelihood decoder for the auxiliary channel $Q_{Y|X}$. Given $\mathcal{P}_M$ and an input symbol sequence of length $N$, we use a distribution approximation based on the variational distance minimization~\cite[Sec. 2.5.2]{CIT-111} to find $q_m$ such that $q_mN \geq 1$. From a sequence $\bm{x}=[x_1,x_2,\ldots,x_n]$ sampled with probabilities $\mathcal{Q}_M$ and corresponding channel outputs $\bm{y}=[y_1,y_2,\ldots,y_n]$,~\eqref{eq:MI_w} can be numerically estimated as

\begin{equation}
\begin{aligned}
\widetilde{\text{MI}}_{\text{num}} \coloneqq & -\frac{1}{N}\sum_{{m=1}}^M\sum_{{x_n\in \mathcal{B}_m}}w_m\log_2\left(\frac{|\mathcal{B}_m|}{N}\right)\\
&-\frac{1}{N}\sum_{{m=1}}^M\sum_{{x_n\in \mathcal{B}_m}}w_m \log_2(w_m)\\
&+\frac{1}{N}\sum_{{m=1}}^M\sum_{{x_n\in \mathcal{B}_m},y_n}w_m \log_2\left(Q_{X|Y}(x_n|y_n) \right),    
\end{aligned}    
\label{eq:MI_cost}
\end{equation}
where $\mathcal{B}_m = \{n:x_n = \tilde{c}_m\}$. Then $q_m = |\mathcal{B}_m|/N$ where $|\cdot|$ is the cardinality of the set. Evaluating~\eqref{eq:MI_cost} for the distribution $\mathcal{P}_M$ from a sequence sampled according to the distribution $\mathcal{Q}_M$ is an instance of importance sampling. 

 Our goal is to maximize~\eqref{eq:MI_cost} by finding optimal values of $p_m$. We carry out the optimization with nested loops, summarized in Alg.~\ref{alg:wlearning}. In the beginning of the outer loop, we calculate the approximation $\mathcal{Q}_M$ from $\mathcal{P}_M$ using variational distance minimization~\cite[Sec. 2.5.2]{CIT-111} and initialize the weights $\mathcal{W}_M$. In the inner loop, we calculate $\tilde{\mathcal{P}}_M$ from $\mathcal{W}_M$ based on the fixed $\mathcal{Q}_M$. We use $\tilde{\mathcal{P}}_M$ for normalizing the constellation points, evaluate~\eqref{eq:MI_cost} and update $\mathcal{W}_M$ using gradient ascent. Gradients of~\eqref{eq:MI_cost} with respect to the weights $w_m$ required for the gradient ascent can be calculated using automatic differentiation. In the end of the outer loop, we calculate $\mathcal{P}_M$ from $\mathcal{Q}_M$ and the updated $\mathcal{W}_M$. As it is common in machine learning, we will refer to the outer loop as an epoch and the inner loop as a batch. We denote the batch size with $B$ and the total number of epochs with $E$. 

To see that the above procedure indirectly makes use of the exact gradients of~\eqref{eq:MI_cost} in terms of the constellation point probabilities $\mathcal{P}_M$, we take partial derivatives of~\eqref{eq:MI_cost} with respect to $w_m$ to have
 \begin{minipage}{.465\textwidth}
    \begin{algorithm}[H]
\caption{Weight learning method for MI maximization}
\label{alg:wlearning}
\begin{algorithmic}
\Require $\mathcal{C}_M$, $\mathcal{P}_M$ 

\For{epoch $= 1$ to $E$}
    \State $\mathcal{Q}_M \gets $Approximate($\mathcal{P}_M$)
    \State $\mathcal{W}_M=\mathcal{P}_M/\mathcal{Q}_M$
    \For{batch $= 1$ to $B$}
        \State $\tilde{\mathcal{P}}_M = \mathcal{W}_M\cdot \mathcal{Q}_M$
        \State $\tilde{\mathcal{C}}_M \gets $Normalize$(\mathcal{C}_M,\tilde{\mathcal{P}}_M)$
        \State Draw $\bm{x}$ from $\tilde{\mathcal{C}}_M$ according to $\mathcal{Q}_M$
        \State Compute~\eqref{eq:MI_cost} from $\bm{x}$ and channel outputs $\bm{y}$
        \State Compute gradients
        \State Update $\mathcal{W}_M$ using gradient ascent
    \EndFor
    \State $\mathcal{P}_M = \mathcal{W}_M\cdot \mathcal{Q}_M$
\EndFor

\State \textbf{return} $\mathcal{P}_M$
\end{algorithmic}
\end{algorithm}
\end{minipage}

\begin{equation}
\begin{aligned}
  \frac{\partial \widetilde{\text{MI}}_{\text{num}}}{\partial w_m} =& -\frac{|\mathcal{B}_m|}{N}\log_2\left(\frac{|\mathcal{B}_m|}{N}\right)\\
  &-\frac{|\mathcal{B}_m|}{N}\log_2(w_m)-\frac{|\mathcal{B}_m|}{N}\log_2(\mathrm{e})\\
  &+\frac{1}{N}\sum_{{x_n\in \mathcal{B}_m,y_n}}\log_2\left(Q_{X|Y}(x_n|y_n) \right)\\
  &+\frac{1}{N}\sum_{j=1}^M\sum_{{x_n\in B_j,y_n}}w_j \frac{\partial}{\partial w_m}\log_2\left(Q_{X|Y}(x_n|y_n) \right).
\end{aligned}    
\end{equation}
Considering that $q_m = {|\mathcal{B}_m|}/{N}$, $w_m={p_m}/{q_m}$ and using ${{\partial f}/{\partial w_m}=q_m{\partial f}/{\partial p_m}}$, we can show that

\begin{equation}
\begin{aligned}
  \frac{\partial \widetilde{\text{MI}}_{\text{num}}}{\partial p_m} =& -\log_2(p_m)-\log_2(\mathrm{e})\\
  &+\frac{1}{|\mathcal{B}_m|}\sum_{{x_n\in \mathcal{B}_m,y_n}}\log_2\left(Q_{X|Y}(x_n|y_n) \right)\\
  &+\sum_{j=1}^M  \frac{p_j}{|B_j|}\sum_{{x_n\in B_j,y_n}}\frac{\partial}{\partial p_m}\log_2\left(Q_{X|Y}(x_n|y_n) \right).
\end{aligned}    
\label{eq:MI_num_grad}
\end{equation}
The terms in~\eqref{eq:MI_num_grad} are the same as those in \cite[(3)]{Aref22}. Note that $w_m$ does not affect the statistics of the batches, therefore we can optimize with the true gradient computed directly using automatic differentiation, which allows for straightforward implementation. This is in contrast to \cite{Aref22}, where additional gradient terms must be computed and added to the gradient computed via automatic differentiation.

Under bit-metric decoding, the AIR is given by the BMI. We can optimize for the BMI similarly using
\begin{equation}
\begin{aligned}
\widetilde{\text{BMI}}_{\text{num}} =& \frac{-1}{N}\sum_{{m=1}}^M\sum_{{x_n\in \mathcal{B}_m}}\mkern-15mu\left(w_m\log_2\left(\frac{|\mathcal{B}_m|}{N}\right) \mkern-4mu + \mkern-4muw_m \log_2(w_m)\right)\\
&+\frac{1}{N}\sum_{{m=1}}^M\sum_{{x_n\in \mathcal{B}_m},y_n}w_m\sum_{k=1}^K \log_2\left(Q_{B^k|Y}(b_n^k|y_n) \right),    
\end{aligned}    
\label{eq:BMI_cost}
\end{equation}

where $K=\log_2(M)$ and $Q_{B^k|Y}$ is the posterior probability that the $k$th bit $b_n^k\in \{0,1\}$ of the $n$th symbol $x_n$ was transmitted given the received symbol $y_n$, assuming that bit labels of length $K$ are mapped to each symbol.

A related algorithm was proposed in~\cite[Sec.~II]{Aoudia2020JointLO}, where the inner expectation is computed explicitly, eliminating the need for a trainable sampling mechanism by using samples drawn independently of the constellation point probabilities. The algorithm in~\cite[Sec.~II]{Aoudia2020JointLO} always samples constellation points according to a fixed distribution, e.g., the uniform distribution, regardless of the learned constellation probabilities. However, such a training is only suitable for channels where the noise and impairments are independent of the constellation points. For generic channels like optical nonlinear channels, the noise statistics depends on the sequence of transmitted symbols, hence it becomes important to sample the transmit symbols from the learned constellation. Our proposed Alg. 1 adapts trainable sampling by drawing symbols from a fixed distribution $\mathcal{Q}_M$ during each epoch and updating weights $\mathcal{W}_M$ to infer the symbol probabilities $\mathcal{P}_M$. Unlike~\cite{Aoudia2020JointLO}, however, $\mathcal{Q}_M$ is updated at the end of each epoch based on $\mathcal{P}_M$, to keep always, $\mathcal{P}_M\approx\mathcal{Q}_M$ during the whole training process.

\section{Simulation Results and Conclusions}
We apply our proposed method to optimize the constellation point probabilities of a 256-QAM constellation for the AWGN channel under an average power constraint and an 8-PAM constellation for a simplified model of the intensity-modulation direct-detection (IM/DD) channel. 

\subsection{AWGN channel}
For AWGN channels, the family of Maxwell-Boltzmann (MB) distributions is very close to the optimum distribution~\cite{Boecherer2017,Kschischang1993,Delsad2023}. We employ Gaussian demappers exactly matched to the channels. This allows us to eliminate all factors other than the learning method. For other more complex channels, neural-network based demappers can be employed. We learn using the COCOB optimizer~\cite{Orabona2017} for $200$ epochs with $20$ batches per epoch and a batch size of $2^{15}$. We fix the constellation points $\mathcal{C}_M$ to the conventional 256-QAM constellation with Gray bit-labeling and initialize their probabilities $\mathcal{P}_M$ to a strongly shaped MB distribution. As with other gradient-based methods, our optimization algorithm is susceptible to getting trapped in local minima. Therefore, when the optimal distribution is unknown for a given channel, it is crucial to initiate the optimization from multiple random distributions to increase the likelihood of finding the global optimum. We validate the optimized probabilities using MI and BMI estimates via 2D Gauss-Hermite quadrature~\cite{Jckel2005ANO}.

\begin{figure}
    \centering
    \includegraphics[width=\columnwidth]{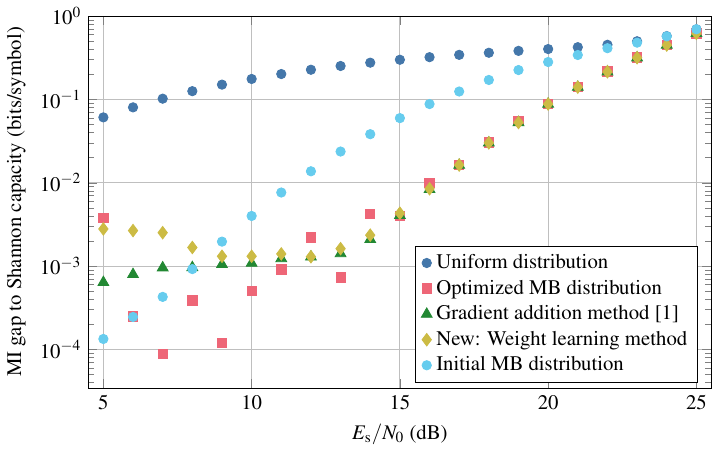}
    \caption{Gap of MI to Shannon capacity for uniform and PCS 256-QAM. The symbol probabilities have been optimized to maximize MI for the AWGN channel under the average power constraint.}
    \label{fig:AWGN_MI}
\end{figure}

\begin{figure}
    \centering
    \includegraphics[width=\columnwidth]{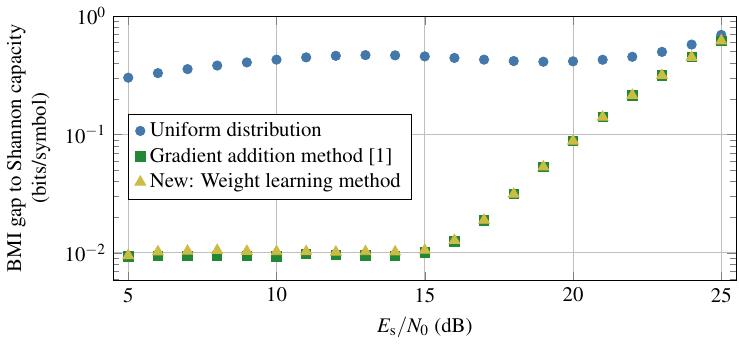} 
    \caption{Gap of BMI to Shannon capacity for uniform and PCS 256-QAM. The symbol probabilities have been optimized to maximize BMI for the AWGN channel under the average power constraint.}
    \label{fig:AWGN_BMI}
\end{figure}

Fig.~\ref{fig:AWGN_MI} shows the MI gap to Shannon capacity ($\log_2(1+E_\text{s}/N_0)$), while Fig.~\ref{fig:AWGN_BMI} shows the BMI gap. Initial Maxwell–Boltzmann (MB) distributions perform well at low $E_\text{s}/N_0$, but diverge from optimized MB distributions at higher $E_\text{s}/N_0$. The proposed weight learning method closely matches both the gradient addition approach from~\cite{Aref22} and optimized MB results. Across all $E_\text{s}/N_0$, MI differences between methods remain below 0.01 bits/symbol—on the order of numerical accuracy for sample based estimation. At low SNRs, PCS-shaped 256-QAM constellations assign very low probabilities to outer points, requiring longer training and careful hyperparameter tuning for convergence. While~\eqref{eq:MI_cost} and~\eqref{eq:BMI_cost} are related, their optimal distributions may differ. PCS substantially reduces the BMI gap, whereas MI gains over the uniform baseline are more modest. As $E_\text{s}/N_0$ decreases, the BMI gap saturates around $10^{-2}$, while the MI gap from our method increases slightly. These discrepancies are minor but could be further reduced by tuning learning parameters such as $B$ and $E$ in Alg.~\ref{alg:wlearning}.


A study of the proposed method under a peak power constraint on the AWGN channel showed excellent agreement with the Blahut–Arimoto algorithm in both mutual information and constellation point probabilities. These results are omitted due to space constraints, but the corresponding code and all results from this work are available at \cite{github}.

\subsection{IM/DD channel}
For the IM/DD channel, we use the following simplified model:
\begin{equation}
    y=|x+n_1|^2+n_2,
    \label{eq:IMDD}
\end{equation}
where $n_1\sim \mathcal{N}(0,\sigma_1^2)$ and $n_2\sim \mathcal{N}(0,\sigma_2^2)$. The term $n_1$ approximates the inter-symbol interference (ISI) seen in a practical IM/DD system and $n_2$ accounts for the detection and electrical amplification noise. Unlike the AWGN channel, the optimal demapper for the IM/DD channel \eqref{eq:IMDD} does not have a closed-form expression. The characteristic function $\varphi_{Y|X}(t)$ of the channel law $p_{Y|X}$ can however be written in closed form as:
\begin{equation}
    \varphi_{Y|X}(t) = \frac{1}{\sigma_1^2 \sqrt{1-2\mathrm{j}\sigma_1^2 t}}\exp\left(\frac{\mathrm{j}x^2t}{1-2\mathrm{j}\sigma_1^2 t} - \frac{\sigma_2^2 t^2}{2}\right)
    \label{eq:CF_IMDD}
\end{equation}
where $j=\sqrt{1}$. The density $p_{Y|X}(y)$ for a given $y$ and $x$ can be computed numerically from the inverse Fourier transform of the characteristic function \eqref{eq:CF_IMDD}. As with the AWGN channel, using demappers matched exactly to the channel isolates the learning method as the sole factor in our comparisons.

\begin{figure}
    \centering
    \includegraphics[width=\columnwidth]{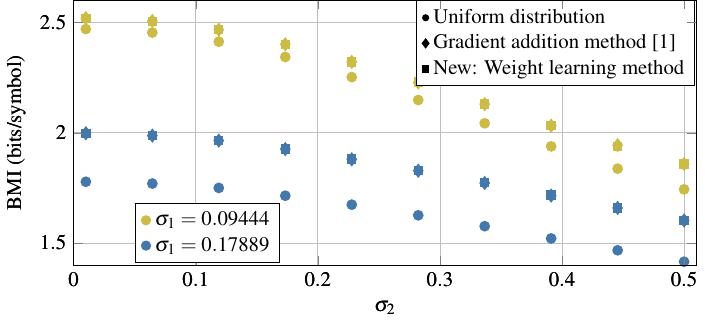}
    \caption{BMI for uniform and PCS 8-PAM. The symbol probabilities have been optimized to maximize BMI for the IM/DD channel.}
    \label{fig:IMDD_BMI}
\end{figure}

\begin{figure}
    \centering
    \includegraphics[width=\columnwidth]{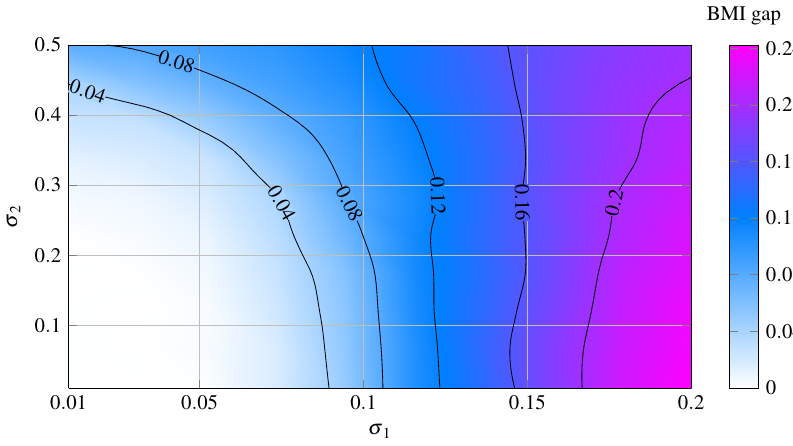}
    \caption{Contour map of the BMI gap between PCS and uniform 8-PAM for the IM/DD channel over the noise parameters $\sigma_1$ and $\sigma_2$.}
    \label{fig:IMDD_mesh}
\end{figure}

The constellation $\mathcal{C}_M=\{\sqrt{c}\,:\,c=0.1+i,\,i= 0,1,\ldots,M \}$ with Gray bit-labeling is initialized with uniform probabilities. Fig.\ref{fig:IMDD_BMI} shows the optimized BMI alongside the BMI for a uniform distribution, for two fixed values of $\sigma_1$ and varying $\sigma_2$. The proposed weight learning method performs comparably to the gradient addition method. A larger gain from PCS is observed when $\sigma_1$ is high, corresponding to stronger ISI. Fig.\ref{fig:IMDD_mesh} presents a contour plot of the BMI gain of PCS over uniform 8-PAM across $\sigma_1$ and $\sigma_2$ values. Black isolines indicate constant BMI gap levels, highlighting the impact of $\sigma_1$ versus $\sigma_2$. The irregular isoline shapes reflect the nonlinear characteristics of the IM/DD channel.

\section{Conclusion}
In this work, we propose cost functions with a corresponding weight learning method that allows for efficient optimization of either MI or BMI using standard automatic differentiation coupled with gradient ascent. We show analytically that our new approach also uses exact gradients of the constellation point probabilities for the optimization as in~\cite{Aref22}. The simulation results of the proposed weight learning method closely match the results from the method in~\cite{Aref22} for the AWGN channel and a simplified model of the optical IM/DD channel. The proposed cost functions are easier to implement as they do not require the manual computation of additional gradient terms and are also applicable to other channel models.
\bibliographystyle{IEEEtran}
\bibliography{main}

\end{document}